\documentclass{article}
\RequirePackage[numbers,compress]{natbib}

\usepackage[accepted]{icml2024}

\usepackage[utf8]{inputenc}   
\usepackage[T1]{fontenc}      
\usepackage{url}              
\usepackage{booktabs}         
\usepackage{amsfonts}         
\usepackage{nicefrac}         
\usepackage{microtype}        
\usepackage{xcolor}           
\usepackage[pdftex]{graphicx} 
\usepackage{siunitx}          
\usepackage[]{algorithm2e}
\RestyleAlgo{ruled}
\definecolor{darkgreen}{rgb}{0,0.75,0}
\definecolor{darkyellow}{rgb}{0.7,0.7,0.2}

\newcommand\tensor[1]{\ensuremath{\mathcal{#1}}}
\newcommand\tset[1]{\ensuremath{\{#1\}}}
\def\this{FLAASH}

\icmltitlerunning{\this: Flexible Accelerator Architecture for Sparse High-Order Tensor Contraction}

\begin{document}

\twocolumn[
\icmltitle{\this: Flexible Accelerator Architecture \\
for Sparse High-Order Tensor Contraction}

\begin{icmlauthorlist}
    \icmlauthor{Gabriel Kulp}{osu}
    \icmlauthor{Andrew Ensinger}{osu}
    \icmlauthor{Lizhong Chen}{osu}
\end{icmlauthorlist}

\icmlaffiliation{osu}{School of EECS, Oregon State University, Corvallis, OR}

\icmlcorrespondingauthor{Gabriel Kulp}{kulpga@oregonstate.edu}
\icmlcorrespondingauthor{Lizhong Chen}{chenliz@oregonstate.edu}

\icmlkeywords{Tensor Contraction, Sparsity, Machine Learning, ICML}

\vskip 0.3in
]

\printAffiliationsAndNotice{}

\begin{abstract}
  Tensors play a vital role in machine learning (ML) and often exhibit properties best explored while maintaining high-order.
Efficiently performing ML computations requires taking advantage of sparsity, but generalized hardware support is challenging.
This paper introduces \this{}, a flexible and modular accelerator design for sparse tensor contraction that achieves over $25\times$ speedup for a deep learning workload. 
Our architecture performs sparse high-order tensor contraction by distributing sparse dot products, or portions thereof, to numerous Sparse Dot Product Engines (SDPEs).
Memory structure and job distribution can be customized, and we demonstrate a simple approach as a proof of concept.
We address the challenges associated with control flow to navigate data structures, high-order representation, and high-sparsity handling.
The effectiveness of our approach is demonstrated through various evaluations, showcasing significant speedup as sparsity and order increase.

\end{abstract}

\section{Introduction}
Tensors are ubiquitous in machine learning, with deep learning models commonly using high-order tensors to represent data, neural activations, and parameters.
Sparse tensor computations can increase the efficiency of processing large-scale data.
Sparsity presents an opportunity to compress data in storage and skip superfluous computations.
However, software approaches have limited ability to harness the parallel nature of the arithmetic, and hardware support remains challenging.

Enhancing the efficiency of \emph{tensor contraction}, an operation that extends matrix multiplication to higher dimensions, could accelerate various types of machine learning \cite{SparseAccelSurvey}. 
Neural network layers can be replaced with a single tensor contraction or tensor decomposition (decomposition requires many contraction operations when optimizing and evaluating) \cite{TensorContractionLayers,TensorRegressionNetworks}.
Additionally, \citet{SupervisedLearningQuantumInspired} note that tensor contraction can be used directly as a method of machine learning.
Currently, these operations are expensive compared to the standard highly-optimized operations on dense tensors of activations and weights.

There is a pressing need to improve hardware support for high-order tensors with unstructured sparsity.
Recent advancements have made progress in addressing challenges in tensor contraction in general, but they do not take on this specific problem.
\citet{OpSparse} focus only on GPU implementations.
\citet{BarrierFreeLargeScaleSparse} target sparse convolution operations, and \citet{AcceleratingSparseDeep} deal with structured sparsity.
It is common for existing work to concentrate on only matrices \cite{OpSparse,AcceleratingSparseDeep,SpArch,ExtendingSparseTensor,DualsideSparseTensor,outerspace}, or on sparse-dense operations\cite{tensaurus}.
Performance improvements from our work will directly enable and enhance the improvements promised in these approaches (e.g., reducing the parameter count by more than 65\% while maintaining classifier performance).

Although structured sparsity can be exploited \cite{AcceleratingSparseDeep}, imposing structure requirements can limit the ability to achieve high sparsity levels.
A method for computing on and storing unstructured sparse data is essential.
Unstructured sparsity is more challenging to implement, since it allows no assumptions, while structured sparsity relies on meeting specific conditions.
Typically, sparse tensor hardware focuses on matrices, with higher-order tensors requiring reformatting or posing difficulties in memory access patterns and caching.
Arbitrary-order hardware therefore offers the greatest flexibility.
These requirements are met by \cite{extensor}, though our approach offers greater flexibility by using modularity to easily support alternative tensor storage formats, scheduling strategies, and caching behavior.

To address these challenges, we introduce \this{} (FLexible Accelerator Architecture for Sparse High-order tensor contraction), a groundbreaking approach to sparse tensor contraction architecture. 
Our results demonstrate the significant potential of \this{}.
We evaluate its performance in two categories of benchmarks: synthetic and deep learning.

\newpage
The main contributions of this paper are:
\begin{enumerate}
    \item Proposing a novel architecture for computing tensor contractions with modular flexibility in both the arithmetic decomposition and the data formats.
    \item Implementing reference hardware of the architecture to contract sparse tensors in Compressed Sparse Fiber format.
    \item Evaluating the hardware implementation, via a conservatively simulated implementation, by conducting experiments against a common data science software package for sparse tensor operations. 
\end{enumerate}

\section{Background and Related Work}


A \emph{tensor} is a multidimensional array, used to represent and manipulate multilinear data.
Each dimension of a tensor is called a \emph{mode}, and the number of modes is the tensor's \emph{order}.
A matrix, for example, has order 2 and the two modes are rows and columns.
We denote tensors with calligraphic capital letters, as in \(\tensor{X}\in\mathbb{R}^{I\times J\times K}\) (a tensor with three modes), and use \(x_{ijk}\) to represent the element at coordinates \((i,j,k)\).
A subset of coordinates can also be shown as a list \tset{a} for brevity; for example, if \(\tset{a}=\tset{i,j}\), then we can say \(x_{i,j,k}=x_{\tset{a}k}\).

Tensors have many uses as practical data formats.
For example, a gray-scale image is an order-2 tensor, and a color image is order-3.
But tensors can be used for more than just storage: they are also the way to arrange operands for computations that, as discussed in Section~1, are useful for machine learning.

One such computation, \emph{tensor contraction}, extends the idea of a matrix multiplication into higher orders.
Contraction yields the inner-product of two tensors.
In a matrix multiplication, the operation is 
\begin{equation} 
    \tensor{C}_{ik}=\sum_j \tensor{A}_{ij}\tensor{B}_{jk}
\end{equation}
which is a series of dot products matching up every column of one matrix with every row of the other to form one entry of the result.
The result of a matrix multiplication is another matrix, which has order 2.
More generally, the order of the result of a contraction is \(m+n-2\) where \(m\) and \(n\) are the orders of the input tensors. Just like in matrix multiplication, the lengths of each contributing fiber to a  tensor contraction must be equal.

Fiber is a generalization of the concept of a row or column. In tensor contraction, the \emph{fiber} length of one operand's contraction mode must match the fiber length of the other.
Therefore, more generally, \(\tensor{A}\in\mathbb{R}^{I\times J\times K}\) and \(\tensor{B}\in\mathbb{R}^{X\times Y}\) can contract along dimensions \(J\) and \(Y\) only if \(J\) and \(Y\) have the same length.
This operation yields a result \(\tensor{C}\in\mathbb{R}^{I\times K\times X}\) where the shapes of \tensor{A} and \tensor{B} are initially concatenated (\(I,J,K,X,Y\)) and then the contracting modes are removed, leaving modes \(I,K,X\).
The modes that are not contracted are the ones traversed to find fibers to match up for dot products.
These so-called \emph{free modes} are the same modes  represented in \tensor{C}: \(I,K,X\) in this case.
For clarity, we will only contract along the last mode in this example, so the contraction operation can be written \
\begin{equation} 
    \tensor{C}_{\tset{a}\tset{b}}=\sum_i\tensor{A}_{\tset{a}i}\tensor{B}_{\tset{b}i}
\end{equation} 
where \tset{a} and \tset{b} index the free modes and \(i\) indexes the contraction mode.

\subsection{Sparsity}
\emph{Sparse} tensors have a high proportion of entries that are zeros, with only some small number of non-zero values.
When expressed as a percentage, sparsity is the fraction of the total volume (product of all mode lengths) occupied by nonzero values.
The \textbf{n}umber of \textbf{n}on-\textbf{z}ero values is often abbreviated as NNZ.
These zeros obviate the multiplications and accumulations internal to the contraction operation, meaning that sparsity-unaware storage and compute techniques will ``waste time'' processing zeros that could be skipped.
This goal is at the heart of sparsity optimizations: efficiently skipping the zeros in storage and compute.

Sparsity commonly arises in data used for machine learning for many reasons.
Sparse matrices may represent momentary observations in discrete time, such as readings from a photodetector that usually does not detect any photons.
Sparsity often arises in the storage of graphs using the adjacency matrix format, where each coordinate in a matrix represents a connection between nodes referenced by the row and column.
Since many graphs are far from fully-connected, this leads to many zeros representing no connection between those nodes.
Also, sparsity is often desirable in neural network weights (and therefore activations) themselves to reduce storage size or inference FLOP count (assuming the inference procedure is sparsity-aware)\cite{SparseAccelSurvey}.

In unstructured sparsity, there is potential for significant imbalance in the number of nonzero elements in each fiber.
One job may require iterating through 10000 nonzero elements while the next job has only two.
This puts strain on the scheduling system, particularly if it cannot predict that these scheduled jobs will take very different amounts of time to complete.
This challenge is particularly pressing for GPU-like architectures with a shared instruction pointer across many processing elements since the job that completes early will keep the processing element occupied until the longest-running job in the work group completes.
Our architecture instead allows each processing element to operate independently without adding complexity.

Sparse tensors can be stored in memory and on disk in a variety of formats.
\emph{Compressed sparse fiber (CSF)} format is common because it is an extension of the CSR (row) and CSC (column) formats often used for sparse matrices.
Storing in CSF format requires performing the following procedure.
First, a mode is chosen, and all fibers that go in that direction are enumerated.
Next, for each fiber, coordinate-entry pairs are stored in indexed order with all zero elements omitted.
Then, these coordinate-entry pairs are stored end-to-end in some way that enables later retrieval of a fiber.
Often, the approach is to store the start and end pointers for each fiber in an order that makes traversal easy.

While these patterns are easy for a CPU to parse, a CPU is insufficiently parallel and leaves performance on the table.
This means we need a custom architecture to traverse these structures efficiently.
We implement hardware traversal of compressed data structures to skip element-wise zero checks or seeking: all pointers are calculated from  tensor metadata while tensor in-memory size is dominated by NNZ.

\subsection{Prior work}
Prior work includes \emph{software} approaches for sparse tensor contractions and also includes hardware accelerators for sparse \emph{matrix} operations.
We believe that our \emph{modular hardware} accelerator for unstructured high-order sparse \emph{tensor} contractions is a novel contribution.
Our work is similar to \cite{extensor}, but our approach places a strong focus on the modularity of the architecture, leaving the door open to alternative designs of each component while retaining the overall architecture structure.
We made decisions to use particular tensor storage formats, scheduling strategies, and memory setups, since concrete options must be chosen to perform simulations and benchmarks, but in this paper we present a general flexible and modular framework for the construction of sparse tensor contraction accelerator architectures, while \cite{extensor} presents only a single design.

There is disagreement about the optimal choice of data structure for accessing and storing operands, intermediate results, and the final result of contraction.
In particular, many papers on the subject discuss the use of the compressed sparse row/column/fiber (CSR, CSC, CSF) format\cite{OpSparse,NEST,taco} and a coordinate-keyed dictionary format\cite{Sparta,Athena}.

\citet{DualsideSparseTensor} and \citet{SpArch} explicitly work down to the hardware on sparse tensors, but they only consider matrices and outer-products (recall that contraction is an inner-product).
\citet{ExtendingSparseTensor} discuss hardware design of format converters for working with sparse matrices during multiplication, which play a similar role to the ``tensor memory'' portion of \this{}, discussed in Section~3. Their work could be adapted to serve as an implementation of one modular unit in our full architecture.

Prior work also shows that sparse tensor contraction can play an important role in machine learning. 
Tensor contraction is explored in deep learning as a replacement for fully-connected layers\cite{TensorContractionLayers,SupervisedLearningQuantumInspired}, and as novel layers\cite{TensorRegressionNetworks}.
There is recent work on noticing and promoting sparsity in neural network activation tensors\cite{SGDSparsity,SparseTransformers} and in transformer activations.
%
%
\section{Approach}
We present \this{}: a FLexible Accelerator Architecture for Sparse High-order tensor contraction.
This architecture scales well to the available die area, and it efficiently handles sparsity: computation time is dominated by the number of nonzero values rather than the volume of the tensors.
The design is modular in that each component communicates through an abstracted interface, allowing support for multiple storage formats, cache-aware job scheduling, and optional scratchpad use.

Our design offers a complex control flow in a highly parallelized implementation: each SDPE follows its own state machine to traverse through the sparse dot product, but is small enough that many fit on the die.

Finally, our architecture achieves workload balance across processing elements (Sparse Dot Product Engines, or SDPEs), facilitated by the central job queue.
Even if one job specifies a fiber that is more dense than average, the other jobs are not held back from dispatch or completion.
Our architecture instead allows each SDPE to operate independently without adding complexity.
Unlike the GPU latency hiding approach of context switching, our design leverages a surplus of processing elements such that even a high idle fraction can saturate the memory interface.

\begin{figure}
  \centering
  \includegraphics[width=0.65\linewidth]{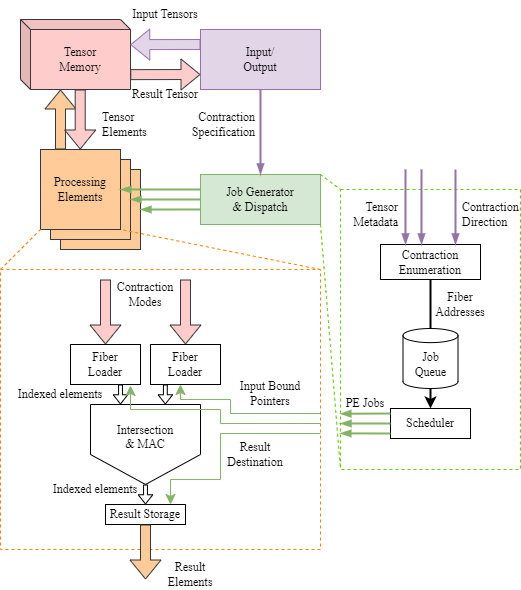}
  \caption{\this\ architecture overview in the center. Detail views below.}
  \label{fig:arch}
\end{figure}
\subsection{Overview}

\begin{algorithm}
\caption{Architecture Level Algorithm}%
 \label{alg:overview}
 \KwData{Tensor \tensor{A}} 
 \KwData{Tensor \tensor{B}}
 \KwResult{Tensor \tensor{C}}

 Save entries of \tensor{A} and \tensor{B} to Tensor Memory\;
 Save pointers of \tensor{A} and \tensor{B} to Job Generator\;
 Calculate Tensor \tensor{C} Metadata\;
$
 \text{Job Count} \gets (\tensor{A}_{\text{Pointer Count}}  - 1) \times (\tensor{B}_{\text{Pointer Count}} - 1)\;
 \text{Jobs Assigned} \gets 0\;
 \text{Jobs Done} \gets 0\;
$

 \While{\text{Jobs Done} < \text{Job Count}}{
  Wait for a PE to be ready\;
  \If{PE has Result}{
    \If{Result Data != 0}{
    Save Result to \tensor{C} memory\;
    Update Pointers and Entry Count\;
    }
    $\text{Jobs Done} \gets \text{Jobs Done} + 1$\; 
    
  }

  \If{\text{Jobs Assigned} < \text{Job Count}}{
    Generate Job fiber bounds\;
    Assign Job to Processing Element\;
    $\text{Jobs Assigned} \gets \text{Jobs Assigned} + 1$\; 
 
  }
 }

 \KwRet{ Pointer to \tensor{C} in memory}\;
\end{algorithm}

\this\ has four high-level components, shown in Fig.~\ref{fig:arch}: processing elements called \emph{Sparse Dot Product Engines (SDPEs)}, job generator \& dispatch, tensor memory, and input/output.
The arithmetic approach is to convert the tensor contraction into a list of dot products, then distribute these dot products over the set of SDPEs.
Each SDPE is a small, simple element that traverses the tensor data structure to process a single dot product.
The SDPE is responsible for gathering all relevant tensor elements, multiplying and accumulating as needed, and requesting that the final accumulator value be written into the result tensor.

The general process this architecture aims to implement is shown in Algorithm~\ref{alg:overview}.
At the top level, the accelerator takes as input two sparse tensors and the modes along which to contract them.
The input/output manager handles any direct memory access requirements or buffer copies to ensure that all required data is available to the tensor memory as needed.
The job generator then reads the metadata of these tensors to generate a list of dot products that will accomplish the contraction.
These dot products (represented as the portions of the input tensors to be read) are then distributed to the SDPEs, which handle the translation from dot product ranges to tensor indices, and request elements from the tensor memory.
The tensor memory then services these requests, responding with tensor elements and their index along the fiber.

The scheduler maintains a queue of jobs and distributes them to SDPEs as they finish each dot product job.
When all jobs are complete and the last result is written, the accelerator signals that it has finished the contraction and returns a pointer to the resulting tensor data structure.
\begin{algorithm}
\caption{Sparse Dot Product Engine (SDPE)}%
 \label{alg:sdpe}
 \KwData{Tensor \tensor{A} Fiber Start and End Pointers} 
 \KwData{Tensor \tensor{B} Fiber Start and End Pointers}
 \KwData{Tensor \tensor{C} Fiber Entry Destination and Index}
 
$
 \tensor{A}_{Pointer} \gets \text{Tensor } \tensor{A} \text{ Fiber Start Pointer}\;
$
$
 \tensor{A}_{End} \gets \text{Tensor } \tensor{A} \text{ Fiber End Pointer}\;
 $
 $
 \tensor{B}_{Pointer} \gets \text{Tensor } \tensor{B} \text{ Fiber Start Pointer}\;
 $
 $
 \tensor{B}_{End} \gets \text{Tensor } \tensor{B} \text{ Fiber End Pointer}\;
$

 \While{$\tensor{A}_{Pointer} < \tensor{A}_{End} \text{ and } \tensor{B}_{Pointer} < \tensor{B}_{End}$}{
    $\text{Entry}\tensor{A} \gets \tensor{A}_{Memory}[\tensor{A}_{Pointer}]\;$
    $\text{Entry}\tensor{B} \gets \tensor{B}_{Memory}[\tensor{B}_{Pointer}]\;$
    $\text{Entry}\tensor{C}_{\text{data}} \gets 0\;$

    \If{$\text{Entry}\tensor{A}_{\text{index}} == \text{Entry}\tensor{B}_{\text{index}}$}{
        $
        \text{Entry}\tensor{C}_{\text{data}} += \text{Entry}\tensor{A}_{\text{data}} \cdot \text{Entry}\tensor{B}_{\text{data}}\;
        \tensor{A}_{Pointer} \gets \tensor{A}_{Pointer} + 1\;
        \tensor{B}_{Pointer} \gets \tensor{B}_{Pointer} + 1\; 
        $
    }
    \ElseIf{$\text{Entry}\tensor{A}_{\text{index}} > \text{Entry}\tensor{B}_{\text{index}}$}{
        $\tensor{B}_{Pointer} \gets \tensor{B}_{Pointer} + 1$\; 
    }
    \Else{
        $\tensor{A}_{Pointer} \gets \tensor{A}_{Pointer} + 1$\; 
    }
 }
 Write \tensor{C} Fiber Entry to Destination\;
\end{algorithm}

\subsection{Sparse Dot Product Engines}

The SDPE design is optimized to be small so that many SDPEs can be tiled in limited die area, and to scale the design to fit the needs of the larger processor design.
One SDPE is comprised of two fiber loader units, an intersection \& MAC unit, a result storage unit, and a local job queue.

An in-depth view of an SDPE can be seen in Figure ~\ref{fig:arch} and the algorithm is shown in Algorithm~\ref{alg:sdpe}.
The \emph{fiber loader units} maintain a local FIFO (first-in-first-out queue) of sequential nonzero tensor elements, represented as pairs of coordinate and value.
Each FIFO represents a portion of one fiber to contract, with the fiber getting progressively fetched and loaded at the top, and entries getting pulled from the bottom to accumulate the dot product.
The fiber loader units are responsible for communicating with the tensor memory to request chunks of the tensor data structure; they traverse their local view of the data structure to know what to fetch next as the dot product progresses.

The \emph{intersection \& MAC unit} pulls from these FIFOs to search for locations where nonzero entries of each fiber align.
When the index within the fiber is the same for both inputs, a collision occurs.
This triggers a multiply-accumulate (MAC) inside the intersection unit, adding to its internal accumulator.
If the index from fiber loader unit \texttt{A} is greater than that of \texttt{B}, then the intersection unit discards the current element from the \texttt{B} fiber and pops the next from that FIFO.
After advancing one fiber, the index comparison is repeated.

The \emph{local job queue} serves as a small buffer, giving the central job queue more freedom to distribute one job per cycle, even if multiple SDPEs are able to start work in the same cycle.
With a local job queue, each SDPE can receive its next job and start immediately when it finishes the current one.

Each fiber loader unit also presents a flag to the intersection unit indicating that it has finished adding new pairs to its FIFO.
When this flag is raised and the queue is empty, the job is done.
This triggers sending the accumulator value to the \emph{storage unit}, which queues until the tensor memory is ready to write the result.
This final queue (another FIFO) allows for greater flexibility in the operation of the tensor memory:
For example, the job generator could recognize that there is enough room to store all results in SDPE result queues, and signal that the tensor memory should prioritize all read operations before accepting write requests.
The memory queue also helps address load imbalance issues, since an SDPE can begin working on its next job immediately, even when the execution time of a job is very short from having few nonzero elements, or even an unexpectedly all-zero fiber.

Note that the interface between a fiber loader unit and memory only returns nonzero tensor elements.
This means that the intersection unit is able to cheaply skip ahead while performing the sparse dot product, since seeking does not involve fetching zeros.
Even if the tensor memory uses a data format that requires querying the operand tensors repeatedly to find nonzero elements, this inefficiency is hidden from the SDPE execution model, freeing the implementation of the tensor memory to privately maintain any metadata useful to efficiently traversing the data structure.


\subsection{Job Generation and Dispatch}
A \emph{job} in this context is defined as a specific dot product represented by the information that the SDPE needs to request input elements and store the result.
In mathematical terms, each SDPE calculates 
\begin{equation}
\tensor{C}_{\tset{a}\tset{b}}=\sum_{i=n}^m\tensor{A}_{\tset{a}i}\tensor{B}_{\tset{b}i}
\end{equation}
where the job specifies \(m\) and \(n\) (the limits of a partial or entire dot product), \tset{a} and \tset{b} (the coordinates along the free modes), and the destination index (where to store the result).
To create a list of these specifications, the job generator iterates through all possible values \tset{a} and \tset{b}; it derives the destination index for \tensor{C} by concatenating these lists and removing the contraction nodes.
This is accomplished in a control flow similar to nested \texttt{for} loops: the inner loops iterate over \tset{a} and the outer loops iterate over \tset{b}, both skipping the contraction modes of \tensor{A} and \tensor{B}. Mathematically: 
\begin{equation}
    \tensor{A}_{\text{Start}} = \text{Job\#}/(\tensor{B}_{\text{Pointer Count}} - 1)
    \label{eq:Astart}
\end{equation}
\begin{equation}
    \tensor{B}_{\text{Start}} = \text{Job\#}\%(\tensor{B}_{\text{Pointer Count}} - 1)
    \label{eq:Bstart}
\end{equation}
\begin{equation}
\text{Job Count} = (\tensor{A}_{\text{Pointer Count}}  - 1) \times (\tensor{B}_{\text{Pointer Count}} - 1)
\label{eq:jobcount}
\end{equation}
Where $\tensor{A}_{\text{Start}}$ is the index of the pointer pointing to the start of fiber \{a\}, and $\tensor{B}_{\text{Start}}$ is the index of the pointer pointing to the start of fiber \{b\}. For example: assume Tensor \tensor{A} has 2 free modes (3 pointers), and Tensor \tensor{B} has 3 free modes (4 pointers). Using equations \ref{eq:Astart} - \ref{eq:jobcount}, the following list of contractions shown in table \ref{contractions} is generated which will cover all possible fibers in the order that they should be added to memory for Tensor \tensor{C}. In this example, the fourth job will contain the pointers pointing to the beginning and end of the second fiber of \tensor{A}, and the pointers pointing to the beginning and end of the first fiber of \tensor{B}. There will be a total of $2\cdot3=6$ jobs for this contraction.

\begin{table}
  \centering
  \caption{Generated Contractions}
  \label{contractions}
  \begin{tabular}{|c|c|c|}
    \hline
    Job \# &Tensor \tensor{A} Fiber & Tensor \tensor{B} Fiber\\
    \hline
     0 & 1 &  1 \\
     1 & 1 &  2 \\
     2 & 1 &  3 \\
     3 & 2 &  1 \\
     4 & 2 &  2 \\
     5 & 2 &  3 \\
    \hline
  \end{tabular}
\end{table}

In our proof-of-concept implementation, the job generator is in charge of translating ranges of tensor indices into pointer boundaries (to mark the start and end of a continuous run in memory) for \(\tensor{A}_{\tset{a}i}\forall i\in I\). Generated jobs are added to a queue to await dispatch to SDPEs.

When all iteration through the free modes of \tensor{A} and \tensor{B} is complete, the job generator waits until the queue is empty, continues to wait until all SDPEs are idle, and then signals that the contraction operation is complete.

Note that dot products can be decomposed into smaller parts. For example:  
\begin{equation} \sum_{i=1}^{100}\tensor{A}_i\tensor{B}_i = \sum_{i=1}^{50}\tensor{A}_i\tensor{B}_i+\sum_{i=51}^{100}\tensor{A}_i\tensor{B}_i.
\end{equation}
Our architecture could easily be adapted to take advantage of dot product decomposition to optimize for cache residency.
This chunking could be implemented in the job generator to maintain simplicity in the SDPEs.

We did not implement this method since our focus was not on optimizing memory structure or cache utilization, but the architecture provides this flexibility.
The tensor memory and the job generator must be in agreement about the storage format of the result tensor.
Focusing on the common case where the result is more dense than the operands, our simple solution is to store the result tensor in a dense CSF format that does not require any prediction, balancing, or dynamic allocation.
We leave it to the driver software to sparsify the result tensor.
This design decision allows more freedom in implementing the job generator: it removes write order dependencies that are required for some sparse storage formats, and it also avoids the dynamic requirements of order-independent COO formats (such as trees and hash tables).

\subsection{Tensor Memory}\label{sec:tensor_memory}



The role of the \emph{tensor memory unit} is to interact with system memory on behalf of the SDPEs, performing the translation from tensor operation to memory access.
Recall that tensor contraction requires computing on pairs of fibers, where every fiber of the first tensor must be paired with every fiber of the second.
As mentioned in the prior section, this carries caching implications.
Furthermore, memory may be virtually segmented into read-only operands and a write-only result.
There is plenty of flexibility to implement this unit to efficiently handle caching with the unique needs of any choice of tensor data structure.
From an integration perspective, it may be helpful to also implement allocation in this unit, and facilitate the import and export of data if a separate memory is used rather than direct memory access (DMA) to the host memory.

The tensor memory performs allocation of tensors for operands and the result by maintaining an internal division between allocated and free memory.
The job generator performs the translation from tensor index to memory pointer, though, which only works because of our choice of data format, which we explicitly chose to lead to the smallest implementation.
This means that the SDPEs make memory requests of the tensor memory with direct physical pointers rather than indices.
This works well because the CSF format is simple and allows for up-front pointer calculations without dynamic pointer chasing or reallocation.
If we had chosen a more complex format like a tree or locality-binned hash table, then the tensor memory would be the only logical place to handle state while traversing the operand structures and assembling the result structure.

This fulfils the adjacency requirement: adjacent elements of a fiber in the contraction mode are also stored physically adjacent in memory.
This storage format also allows the tensor memory to preallocate the entire result tensor, which gives the job generator sufficient context to compute the pointer of each result element before the contraction starts.
Our implementation preallocates a dense result in CSF format, which can be converted to a sparse representation in one pass.
This is suboptimal but simplifies the implementation considerably, and since results tend to be more dense than operands, it is not an invalid assumption.

\subsection{Input/Output Interface}
We envision this accelerator as a small component of a larger processor design, sharing system memory through a DMA interface.
The I/O unit would implement this interface.
In a discrete design, this unit would house the PCIe connection to the host system.
We implemented this portion of the simulator as part of the testbench, since the design of high-speed IO interfaces is out of scope for this project.

\begin{figure*}
  \includegraphics[width=\textwidth]{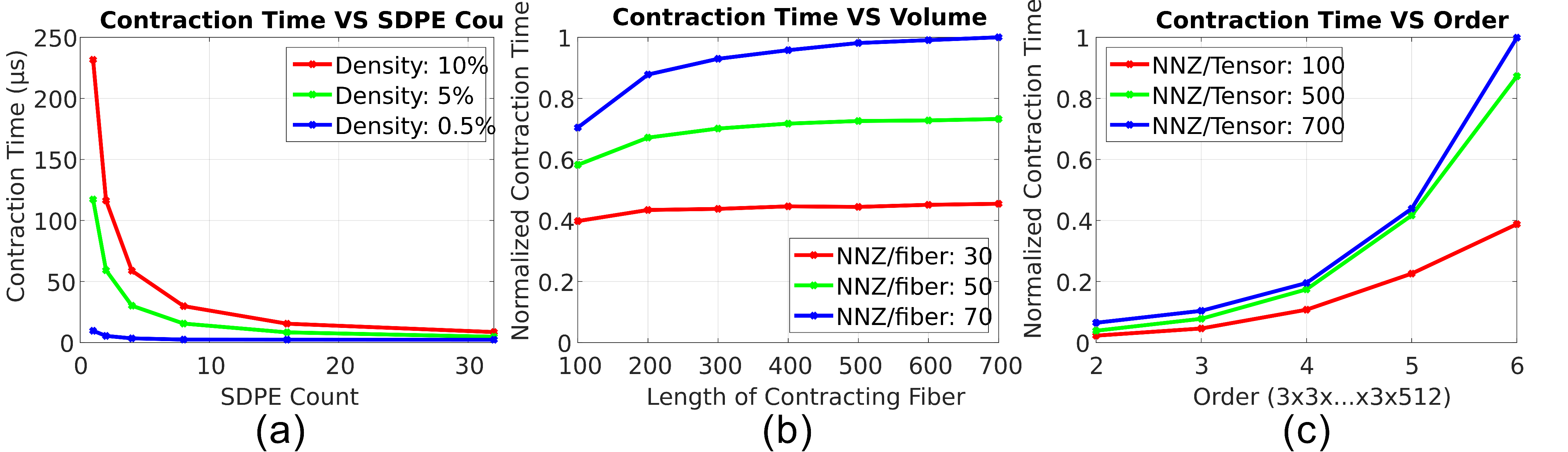}
  \caption{Simulations of Architecture Implemented in Verilog}%
  \label{simulations}
\end{figure*}

\section{Evaluation}

\subsection{Implementation}
We synthesized and simulated a Verilog implementation of our architecture using Xilix Vivado HLS. 
Our simulation assumes a conservative 1GHz clock speed, and only times the contraction operation (not copying operands in or the result out of memory).
This is a reasonable assumption for a small accelerator on a CPU that has direct memory access to main system DRAM.
For comparisons to software alternatives, we evaluate our model against tensor contraction functions found in both the \texttt{Pytorch}\footnote{\url{www.pytorch.org/docs/stable/torch.html}} and \texttt{TensorFlow}\footnote{\url{www.tensorflow.org/resources/libraries-extensions}} libraries, which are widely used in the industry for ML applications.
Software measurements were performed using an 8-core Intel Core i7-11375H (3.3GHz) with 32GB of system RAM. 

\begin{figure*}
  \includegraphics[width=\textwidth]{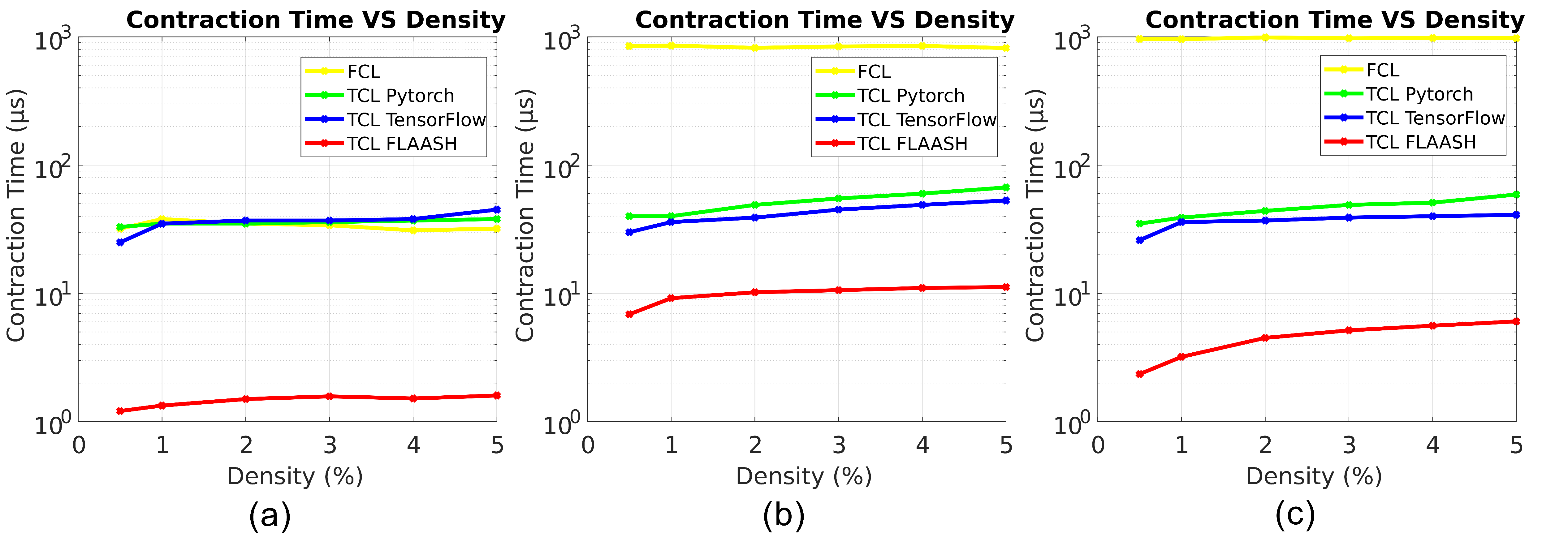}
  \caption{Contraction Times (µs) vs Density (\%) \quad (a) $3 \times 3 \times 1024$ tensor contracted with a $3 \times 1024$ matrix to get an output tensor of $3 \times 3 \times 3$. \quad (b) $7 \times 7 \times 512$ tensor contracted with a $7 \times 512$ matrix to get an output tensor of $7 \times 7 \times 7$. \quad (c) $10 \times 10 \times 100$ tensor contracted with a $10 \times 100$ matrix to get an output tensor of $10 \times 10 \times 10$. Matrices invovled in contraction each have 50\% density.}
  \label{ml_application}
\end{figure*}

Using Intel 16nm technology, the relationship between the total area of the architecture as a function of the number of processing elements used is shown in table \ref{tab:area}. Eight SDPEs were used for the simulation results found in figures \ref{simulations}b,c and \ref{ml_application}a-c.
\newcommand\shape[3]{\ensuremath{(#1\times#2\times#3)}}
\begin{table}[ht]
    \caption{Size of FLAASH Hardware Using Intel 16nm Technology}%
    \label{tab:area}
    \centering
    \begin{tabular}{cc}
        \toprule
        SDPE Count & Total Area $(mm^2)$ \\
        \midrule
        1  & 0.692\\
        2  & 0.703\\
        4  & 0.721\\
        8  & 0.750\\
        16 & 1.938\\
        32 & 3.129\\
        \bottomrule
    \end{tabular}
\end{table}

Tensors are generated randomly, with density as the probability that an individual element will be nonzero. We consider the workloads described by \citet{TensorContractionLayers}: multi-layer perceptrons from AlexNet and VGG-19, replaced with equivalent tensor contractions.
We chose densities to cover a range of empirical results.
For example, \citet{MovementPruningAdaptive} suggest pruned BERT-base encoders achieve around 3\% density, while recurrent neural networks from \citet{PruneNotPrune} are below 10\%.
\citet{SGDSparsity} suggest as low as 6.5\% density, and \citet{SparseTransformers} suggests 3\% to as low as 0.5\% density in transformer activations for some layers.

\subsection{Synthetic Workload}

Fig.~\ref{simulations}a demonstrates the relationship between the contraction time and the number of SDPEs. The tensors contracted in this set of simulations had the shape $7 \times 7 \times 512$. It is valuable to note that as the density decreases, the SDPE count becomes less important because the time it takes to complete a single job is faster than the time it takes to read results and assign jobs to all other SDPEs before looping back. This behavior is a result of our sequential round-robin job distribution strategy. Even at 10\% density there is little improvement with adding more than 32 processing elements. The architecture was given 8 SDPEs for the remaining simulations. 

In Fig.~\ref{simulations}b, we can see that as the Number of Non Zero elements (NNZ) is held constant, there is little difference in contraction time even as the volume of the tensor is increased 7-fold. This is the objective of the architecture: we aim to make the contraction time dependent on the number of non-zero elements in the tensor and not on its dimensions. Only non-zero elements are represented in tensor memory, so the additional volume is only accounted for in the pointers. Note that the contraction time increases proportionally to NNZ in each contracting fiber. The shape of the tensors used in this set of simulations is $5 \times 5 \times n$. 

In Fig.~\ref{simulations}c, the contraction time is compared to the order of the tensor. In this set of simulations, if the contracting tensor is of order N, the first $N-1$ dimensions have length 3 and the contraction dimension is of length 512. These tensors are contracted against a $3 \times 512$ matrix. Note that the contraction time does increase with order because, as the number of pointers increases, the number of jobs does too, but the contraction time will increase at a rate far lower than the rate the volume is increasing ($N^3$) if there is a constant number of non-zero elements.
\subsection{Machine Learning Workload}
Fully Connected Layers (FCLs) typically do not leverage sparsity well and thus consume an unnecessary amount of time because a large number of input nodes are processed to see if they are non-zero. An FCL in a neural network can be replaced by a tensor contraction operation to greatly reduce the number of input parameters \citet{TensorContractionLayers}. This can be done by using a Tensor Contraction Layer (TCL) which is the process of contracting an input tensor $\tensor{T}$ of shape $I_1 \times I_2 \times ... \times I_N$ with a matrix $M$ of shape $I_N \times R_N$ where $R_N < I_N$ to contract the final dimension to a much smaller size. The matrix $M$ can then be adapted through back propagation to learn dimensional trends. In this section, we compare four schemes. The first is a Fully Connected Layer with $I_1 \cdot I_2 \cdot ... \cdot I_N$ input nodes and $I_1 \cdot I_2 \cdot ... \cdot R_N$ output nodes, which will serve as the base case to demonstrate the speedup of using a TCL in ML applications. The second and third schemes are a TCL implemented in software using functions: $\texttt{torch.sparse.mm()}$ and $\texttt{tf.sparse.from\_dense()}$ from two libraries: Pytorch and TensorFlow, respectively. This comparison will demonstrate the CPU performance of executing a TCL and the relative speedup of this method versus an FCL. Finally, the fourth scheme is the implementation of \this{} which will demonstrate the speedup of using specialized hardware to perform tensor contractions.

The  Pytorch and TensorFlow libraries both support dense tensor contraction and sparse matrix contraction, but they do not have functions to perform sparse tensor contraction. An equivalent mathematical operation is to reshape sparse tensors into sparse matrices where the free modes are combined to a single mode. These sparse matrices are then contracted and reshaped to the appropriate dimensions. Note that, for these experiments, only the sparse matrix contraction time is being measured, not the time it takes to reshape. 

As can be seen in Figure~\ref{ml_application}a, when the input tensors have low volume there is no speedup when using a TCL implemented by software, but there is a $23.1\times$ speedup from an average of $\SI{33.7}{\micro\second}$ when using an FCL to $\SI{1.46}{\micro\second}$ when using the FLAASH accelerator. Speedups for each case in figure \ref{ml_application} can be seen in table \ref{tab:DLWorkloads}. There is only \SI{0.375}{\micro\second} (30.6\%) variation in contraction time as density increases from 0.5\% to 5\% for the FLAASH accelerator. For the software tensor contractions there is \SI{5}{\micro\second} (15\%) and \SI{20}{\micro\second} (80\%) variation for Pytorch and TensorFlow contraction functions respectively. 

Figure~\ref{ml_application}b shows the contraction times for an input tensor of size $7 \times 7 \times 512$. The contraction time for the fully connected layer has increased to on average \SI{839}{\micro\second} with no dependence on density. Tensor Contraction times for Pytorch and TensorFlow implementations stay relatively similar to Fig.~\ref{ml_application}a at \SI{52}{\micro\second} and \SI{42}{\micro\second} on average respectively. The contraction times for FLAASH increase to an average of \SI{9.8}{\micro\second} due to the increased NNZ and number of free modes. 

Finally, in Figure~\ref{ml_application}c, the contraction times decrease for all TCL implementations because the NNZ is less compared to Fig.~\ref{ml_application}b but the execution time for the fully connected layer has increased to an average of \SI{975}{\micro\second} because of the increased number of output nodes.

\renewcommand\shape[3]{\ensuremath{(#1\times#2\times#3)}}
\begin{table}[ht]
    \caption{Avg Speedup of \this{} in contrast to other methods}%
    \label{tab:DLWorkloads}
    \centering
    \begin{tabular}{cccc}
        \toprule
        Shape & \textcolor{blue}{\texttt{FCL}} & \textcolor{darkgreen}{\texttt{Pytorch}} & 
        \textcolor{darkyellow}{\texttt{TensorFlow}}  \\
        \midrule
        $(3, 3, 1024)$  & $23.1\times$ & $24.5\times$ & $24.8\times$\\
        $(7, 7, 512)$   & $85.3\times$ & $5.26\times$ & $4.27\times$\\
        $(10, 10, 100)$ & $218\times$ & $10.3\times$ & $8.16\times$\\
        \bottomrule
    \end{tabular}
\end{table}

\section{Conclusion and Future Work}
Sparsity in tensors is crucial for efficient ML computations, but providing hardware support is challenging.
We demonstrate that \this\ performs well even in high-sparsity and high-order contexts, providing a significant speedup to machine learning workloads.
FLAASH opens up possibilities for a variety of future scheduling, cache, and data structure optimizations.
Future work could include comparing different data structures and cache management strategies, and exploring opportunities to decompose dot products into shorter jobs.


\bibliographystyle{icml2024}
\bibliography{citations}

\end{document}